\documentclass{article}

\usepackage{mathrsfs}
\usepackage{graphicx}
\usepackage[section]{placeins}
\usepackage{amssymb}
\usepackage{amsmath}
\usepackage{mathrsfs}
\usepackage[sort&compress,numbers]{natbib}
\usepackage[colorlinks,breaklinks,bookmarks,pdfauthor=
 Martín A. Mosquera,pageanchor,citecolor=blue]{hyperref}

\title{Simple isotherm equations to fit type I adsorption data}

\author{Mart\'in A. Mosquera\\
\small Department of Chemistry\\[-0.8ex]
\small Purdue University, 560 Oval Drive, West Lafayette, IN 47907, USA\\
\small \texttt{mmosquer@purdue.edu}
}

\begin{document}

\newcommand*{\mc}[1]{\mathcal{#1}}
\newcommand*{\ms}[1]{\mathscr{#1}}
\newcommand*{\mb}[1]{\mathbf{#1}}
\newcommand*{\mr}[1]{\mathrm{#1}}

\maketitle

\begin{abstract}
A simple model to fit experimental data of adsorption of gases and vapours
on microporous adsorbents (type I isotherms)
is proposed. The main assumption is that the adsorbate phase can be
divided into identical and non-interacting effective subsystems. 
This gives rise to a simple multiparametric isotherm based on the grand canonical
ensemble statistics, whose functional form is a ratio of two polynomial functions.
The parameters are interpreted as effective equilibrium constants. A simplified isotherm that
reduces the number of adjustable parameters with respect to the general
isotherm is also proposed. We show how to 
use these isotherms to fit the adsorption data in such way that the parameters
have statistical significance.
Due to their high accuracy, both isotherms  can be
used to estimate thermodynamic properties like isosteric and differential
heats of adsorption. A simple method is presented for systems that show
an apparent variation in the coverage limit with temperature. This method avoids
overparametrization and improves fitting deviations. 
Finally, several applications to fitting data, taken from the literature, of adsorption of some gases
on activated carbon, molecular sieving carbon, silica gel, and pillared clays
are presented. 
\end{abstract}

\section{Introduction}
Correlating adsorption data obtained from experiments or computer calculations
is necessary to save time and efforts in additional experimentation or
computing time. Because of the complexity of the equilibrium adsorption
phenomenon, the research of new models of adsorption is still very active. A
good parameter-adjustable model must fit well the experimental data and
predict thermodynamic quantities correctly; therefore, it is necessary to
analyse the model and assess its capabilities. Many industrial applications of
adsorption encompass a wide range of adsorbent saturation. Thus, models
 which offer the experimentalist and
process engineer the possibility to set the number of adjustable parameters are required.

Many adsorbents used in industry and research present certain degree of
heterogeneity in terms of pore size distribution and surface
topography. Depending on each of these factors, different isotherm models are
proposed. For example, lattice gas theories which simplify the structure of the
adsorbent surface by taking into account only the most important adsorption sites
of the adsorbent surface or adsorption pores have been developed. Many of
these theories are consistent with experiments and computer simulations  
\cite{aya2,steele,nikitas,gen,nitta,aranovich,ramirez4,ramirez1,ramirez3,ramirez2,chiang}.
A disadvantage of this kind of models is that many gas-solid systems present
unique characteristics. Thus, they may not be applicable if some assumptions or conditions are violated.

Another commonly used method to obtain isotherms for heterogeneous systems is
the integration over a patchwise topography of adsorption sites \cite{duongg}.
Modified isotherms arise by assuming an adsorption energy distribution
function and an ideal local isotherm like the Langmuir equations. Even though
some of these equations work well for a large number of systems, they present
limitations associated with the assumptions of the particular model, e.g., some models do not
provide the correct Henry's law limit. Moreover, this integration method is somewhat
difficult because the complexity of the adsorption energy distribution makes
it difficult to obtain the analytical expression for the transformed isotherm.
The equations of Sips \cite{sips}, Toth \cite{toth}, and UNILAN \cite{sips2} are the
most widely used isotherms of this type.

In this work simplified isotherms to fit type I adsorption 
data are proposed. The model to derive the isotherms is  the cell model of
adsorption shown by Hill \cite{hill3,hill2,hill1}, which is based on the grand canonical ensemble statistics.
The general isotherm that arises from the cell model has been quite successful to fit experimental
data of adsorption of gases on zeolites \cite{ruth2,rut20,aya1,boden1,boden2,stach,rutfin}.
Recently, we showed how to employ this model to fit isotherms of adsorption of gases and vapours on zeolites \cite{martin} 
and that the parameters could be interpreted as adsorption equilibrium constants. 
Further validation of this generalized statistical thermodynamical adsorption (GSTA) model 
has been shown with excellent results for both adsorption data and thermodynamic properties \cite{mario1,mario2}. 
Here, we present an accurate extension of the GSTA model and a new simplified isotherm is derived from it. 
It is shown that  these isotherm equations can be used to fit type I
experimental adsorption isotherms and predict thermodynamic properties. To illustrate this idea,
 several experimental adsorption isotherms are correctly fitted with these isotherms.
Furthermore, extrapolations to temperatures beyond the experimental temperature
range are possible. Because of the complexity of heterogeneous adsorbents, it
is suggested that the method presented here is semi-empirical and the cells
ensemble can be considered as an effective grand ensemble of
small subsystems. 

\section{Theory}

\subsection{The general adsorption isotherm}
Consider a single-component gas in equilibrium with an adsorbate phase and
suppose that the adsorbate phase can be divided into $M$ identical cells that
are  not interacting in the ensemble picture. 
This means that they are coupled by a very small interaction such that the exchange
of molecules in between cells takes place in a time scale which is large in comparison
with the characteristic time scales of the fundamental molecular processes that occur
 inside each cell. The total number of cells is temperature-independent 
and the cell volume as well. It is plausible to assume 
that a maximum of $n_{\mr{s}}$ molecules can adsorb
on the subsystem, i.e. the adsorbent has a saturation limit.  Under these assumptions
we can express the grand canonical partition function $\Xi$ as a product of 
subsystem's partition functions \cite{hill2}:
\begin{equation}
\Xi= \xi^M
\end{equation}
where 
\begin{equation}
\xi=\displaystyle 1+\sum_{j=1}^{n_{\mr{s}}}K_ja^j
\end{equation}
and $a$ is the activity of the system:
 \begin{equation}
a=\frac{f}{P^0} 
\end{equation}
Since we are concerned with the low pressure regime then $a=P/P^0$. The equilibrium
constant $K_j$ is expressed as
\begin{equation} 
K_j(T)=\exp\Big(\frac{j\mu^0_{\mr{g}}-A_{j}^{\mr{S}}}{k_{\mr{B}} T}\Big)\label{eq:bb}
\end{equation}
where $\mu^0$ is the chemical potential of the gas at the reference pressure $P^0$ and $A_j^{\mr{S}}$
 is a microscopic free energy which can be written in terms
of the partition function of the subsystem with $j$ molecules and volume $v_S$ ($Q_S(T,v_S,j)$):
\begin{equation}
A_j^{\mr{S}}=-k_BT\ln Q_S(T,v_S,j)
\end{equation} 
Now the saturation of the system ($q$) is simply:
\begin{equation}
q = \frac{q_{\mr{m}}}{n_{\mr{s}}}\frac{\displaystyle 
\sum_{j=1}^{n_{\mr{s}}} jK_ja^j}{\displaystyle 1+\sum_{j=1}^{n_{\mr{s}}}K_ja^j}\label{eq:isot}
\end{equation} 
Based on its definition, the term $K_j$ can be regarded as an adsorption equilibrium constant 
\cite{martin}. Therefore, in analogy with chemical equilibrium constants, we can propose
the following relation:
\begin{equation}
\ln K_j=\ln K_j^{\circ} - \frac{\Delta h_j}{RT}
\end{equation} 
where $\ln K_j^{\circ}$ and $\Delta h_j$ are the change of entropy and enthalpy,
respectively. They are related with the adsorption of $j$ molecules onto a
representative microscopic subsystem at the reference pressure $P^0$. In his seminal paper, the Eq.
(\ref{eq:isot}) was also derived by Langmuir \cite{lang2} using chemical kinetics 
 for the case in which a
site can hold several molecules and the adsorbent is composed by non-interacting
sites. In light of his proposal, the term $K_j$ is partitioned as follows:
\begin{equation}
K_j=\frac{1}{j!}\prod_{i=1}^jR_i\label{eq:erres}
\end{equation}
In a similar fashion, $R_j$ obeys the following relation:
\begin{equation}
R_j=R_j^{\circ}\exp\Big(\frac{-\Delta\overline{h}_j}{RT}\Big) 
\end{equation}
where
\begin{eqnarray}
R_j^{\circ} &=& \frac{jK_j^{\circ}}{K_{j-1}^{\circ}}\\
\Delta\overline{h}_j&=& \Delta h_j-\Delta h_{j-1}
\end{eqnarray}
In order to fit an isotherm at constant temperature using Eq. (\ref{eq:isot}), the
adjustable parameters are $\{R_j\}$ instead of $\{K_j\}$. For the case in
which several isotherms at different temperatures have been measured, the
adjustable parameters are $\{\ln R_j^{\circ}\}$, $\{\Delta\overline{h}_j\}$,
and $q_{\mr{m}}$. Hence, a total of $2n_{\mr{s}}+1$ parameters are required.

The present model suggests that $\xi\rightarrow \Xi$ as the subsystem volume
is increased. Thus, in order to predict $q$ we should assume that the maximum
number of molecules $n_{\mr{s}}$ and the volume $v_S$ are very large. 
However if we assume that each $R_j$ is an adjustable parameter, then this would give us a large number
of adjustable parameters with poor statistical confidence \cite{rutfin}. A
solution to this problem is to assume $n_{\mr{s}}$ as a small number of
molecules that adsorb into a microscopic imaginary effective subsystem, and the
parameters $\{R_j\}$ as representative parameters of the experimental
adsorption isotherm. The characteristics of the adsorbate+adsorbent system like molecule-surface and
molecule-molecule interactions are included effectively in these subsystems; probably by a modified
intermolecular potential.
This is ad hoc guess is a convenient picture that allows us to apply the
cell model to complex zeolite and heterogeneous systems in general.

The probability that a subsystem with $j$
adsorbed molecules will be found is (assuming ideal gas phase):
\begin{equation}
p_j=\frac{K_j a^j}{\xi}
\end{equation}
Eq. (\ref{eq:isot}) can be written as:
\begin{equation}
\theta =
\frac{1}{n_{\mr{s}}}\sum_{j=1}^{n_{\mr{s}}}jp_j=
\frac{1}{n_{\mr{s}}}\sum_{j=1}^{n_{\mr{s}}}f_j\label{eq:efes}
\end{equation}
here, $\theta=q/q_{\mr{m}}$, and $f_j$ is a fraction of the $n_s$ molecules that are
found in subsystems with $j$ molecules. At low pressures
the leading term in Eq. (\ref{eq:efes}) is $f_1$ (as stated by the Henry's law),
and, as the pressure increases, the leading is $f_2$, and so forth. Consequently, within
certain pressure range, there is an $f_j$ term that significantly contributes to the
fractional coverage. Therefore, each parameter $K_j$ can be associated with data 
taken in a certain range of the saturation $q$. So we infer that these parameters might have large uncertainties
if there is not enough data in their corresponding intervals  .

The general isotherm (Eq. (\ref{eq:isot})) and its simplifications have been
useful to fit isotherms of adsorption of gases and vapours on zeolites
at different temperatures, where the experimental isotherms show
clearly a saturation limit, or are reported within the same range of
saturation (see Refs. \cite{ruth2,rut20,aya1,boden1,boden2,stach,martin,mario1,mario2}). However, a
great number of adsorption systems do not show this behaviour because
experiments typically are performed along the same pressure range; therefore,
the isotherms apparently show that the saturation limit depends on temperature
(i.e., $q_{\mr{m}}$ tend to increase with a decrease in temperature). Now,
consider certain isotherm at temperature $T_1$, if we consider that
$n_{\mr{s}}=3$, then we have the following adjustable parameters:
$K_1,\,K_2,\,K_3$, and $q_{\mr{m}}(T_1)$. Also, suppose that we
have a second isotherm at temperature $T_2$ and its apparent saturation limit
is $q_{\mr{m}}(T_2)$; if $T_2<T_1$, then $q_{\mr{m}}(T_2)>q_{\mr{m}}(T_1)$.
This would suggest that, as the temperature decreases, new subsystems are
created, but this would be an inconsistency for our model. To solve this problem, we
introduce a new equilibrium parameter $K_{n_{\mr{s}}+1}$ that takes into
account the adsorption of an extra molecule on the subsystem at high pressures
and low temperatures. Because this new parameter describes the adsorption at
these conditions, it is possible that there is not enough experimental data to
estimate $\ln R_{n_{\mr{s}}+1}^{\circ}$ and
$\Delta\overline{h}_{n_{\mr{s}}+1}$. To solve this problem we assume that
$R_{n_{\mr{s}}+1}=R_{n_{\mr{s}}}$ and write
\begin{equation}
K_{n_{\mr{s}}+1}=\frac{R_{n_{\mr{s}}}}{n_{\mr{s}}+1}K_{n_{\mr{s}}}\label{eq:crr1}
\end{equation}
Here we have assumed that the change of both entropy and enthalpy of
adsorption of the $(n_{\mr{s}}+1)$th molecule on the microscopic cell is the
same for the adsorption of the $n_{\mr{s}}$th molecule
 (i.e., $R\ln R_{n_{\mr{s}}+1}^{\circ}= R\ln R_{n_{\mr{s}}}^{\circ}$ and 
 $\Delta\overline{h}_{n_{\mr{s}}+1}=\Delta\overline{h}_{n_{\mr{s}}}$).
 If we suppose this is applicable for $l$ extra molecules, then this result is generalized as follows:
\begin{equation}
K_{n_{\mr{s}}+l}=\frac{R_{n_{\mr{s}}}}{n_{\mr{s}}+l}K_{n_{\mr{s}}+l-1}\label{eq:crr2}
\end{equation}
Now, $\xi$ becomes:
\[
\xi=1+\sum_{j=1}^{n_{\mr{s}}+l}K_ja^j
\]
and the isotherm is:
\begin{equation}\label{eq:isott}
q=\frac{q_{\mr{m}}}{n_{\mr{s}}+l}\frac{\displaystyle \sum_{j=1}^{n_{\mr{s}}+l}jK_ja^j}
{\displaystyle 1+\sum_{j=1}^{n_{\mr{s}}+l}K_ja^j}
\end{equation}
This is our extension to the GSTA model and we will show in section \ref{sec:diss} how to apply this Eq. to obtain
statistically reliable estimation of the parameters. 

\subsection{A simplified adsorption isotherm}
In order to reduce the number of adjustable parameters and improve the confidence
intervals around the parameters' estimates, suppose there is a reference temperature $T^{\circ}$
at which the molecules in the subsystem behave as a microgas. This microgas can be either
two or three dimensional, it depends on the adsorbent characteristics. Hence we obtain:
\begin{equation}
K_j^{\circ}=\frac{(K^{\circ})^j}{j!}
\end{equation}
Under these assumptions, it is obtained the adsorption isotherm:
\begin{equation} 
q=\frac{q_{\mr{m}}}{n_{\mr{s}}+l}\frac{\displaystyle 
\sum_{j=1}^{n_{\mr{s}}+l} \frac{\displaystyle \exp(-\Delta
h_j/RT)}{\displaystyle (j-1)!}(K^{\circ}a)^j}{\displaystyle
1+\sum_{j=1}^{n_{\mr{s}}+l}\frac{\displaystyle \exp(-\Delta
h_j/RT)}{\displaystyle j!}(K^{\circ}a)^j}\label{eq:isomix}
\end{equation}
comparing with Eq. (\ref{eq:isott}) we have:
\begin{equation}
K_j=\frac{(K^{\circ})^j}{j!}\exp(-\Delta h_j/RT)\qquad j=1,\ldots,n_{\mr{s}}
\end{equation}
From Eq. (\ref{eq:crr2}) we obtain:
\begin{equation}
K_{n_{\mr{s}}+l}=\frac{\displaystyle K^{\circ}\exp[-(\Delta h_{n_{\mr{s}}}-
\Delta h_{n_{\mr{s}}-1})/RT]}{\displaystyle n_{\mr{s}}+l}K_{n_{\mr{s}}+l-1}
\label{eq:icr1}
\end{equation}
Here, the fitting parameters are $\{\Delta h_j\}$, $K^{\circ}$, and 
$q_{\mr{m}}$. Hence, we have reduced the number of parameters almost by half. \\

\subsection{Isosteric heat of adsorption}
From Eq. (\ref{eq:bb}) we get
\begin{equation}
\Delta h_j=u_j-jh_{\mr{g}}^0
\end{equation}
where $u_j$ and $h_{\mr{g}}^0$ are the energy of $j$ molecules in a subsystem
and the molar enthalpy of the gas phase at $P^0$, respectively, and $\Delta h_j$
is the change of enthalpy related to the adsorption of $j$ molecules on a
subsystem. In principle $\Delta h_j$ is
temperature-dependent, but for the sake of simplicity, we have assumed that
both $h_{\mr{g}}^0$ and $u_j$ are temperature-independent. The isosteric heat
of adsorption can be calculated by means of the following
formula \cite{surarea}:
\begin{equation}
q^{\mr{st}}=h_{\mr{g}}-u_{\mr{a}}
\end{equation}
where $h_{\mr{g}}$ and $u_{\mr{a}}$ are the molar enthalpy of the gas phase
and adsorbate molar internal energy respectively. Define the following
average for an arbitrary discrete real function $g_j$: 
\[
\langle g_j\rangle_{0}=\sum_j p_j g_j
\]
It is easy to show that the isosteric heat of adsorption is a ratio of covariances:
\begin{equation}
-q^{\mr{st}}=\frac{\langle j\Delta h_j \rangle_{0}-n_{\mr{s}}\theta\langle
\Delta h_j\rangle_{0}}{\langle j^2\rangle_{0}-
n_{\mr{s}}^2\theta^2}\label{eq:isosh}
\end{equation}
where $\theta=q/q_{\mr{m}}$. From the above Eq., the two following limits arise:
\begin{eqnarray}
\lim_{\theta\rightarrow 0}q^{\mr{st}}&=&-\Delta\overline{h}_1\label{eq:hen}\\
 \lim_{\theta\rightarrow 1}q^{\mr{st}}&=&-\Delta\overline{h}_{n_{\mr{s}}}
\end{eqnarray}
This result establishes a physical interpretation to the first and the last enthalpies of molecular 
addition. 

\section{Results and discussion}\label{sec:diss}

The unweighted least-squares
Levenberg-Marquardt algorithm was employed to fit the published experimental data used in this work. 
Details of the results of fitting Eqs. (\ref{eq:isomix}) and (\ref{eq:isott}) to several experimental adsorption data are
shown in Tables 1 and 2; the parameters are tabulated with their corresponding
marginal confidence intervals \cite{bates,seber}. The study of these 
intervals is necessary to avoid over-interpretation of the
parameters \cite{kinn} and to estimate uncertainties in predicted thermodynamic
properties. The parameter $\ln K^{\circ}$ (or $\ln K_j^{\circ}$) were used
instead of $K^{\circ}$ (or $K_j^{\circ}$) due to the restriction $K^{\circ}>0$
(for all $j$). The following deviation parameter was employed to analyse the fittings accuracy: 
\begin{equation} D=\frac{100\%}{N_{\mr{T}}}
\times
\sum_{j=1}^{N_{\mr{temp}}}\sum_{i=1}^{N_{\mr{P}}(j)}\Bigg|\frac{q(P_i,T_j)-q_{\mr{exp}}(P_i,T_j)}
{q_{\mr{exp}}(P_i,T_j)}\Bigg| 
\end{equation} 
where $N_{\mr{temp}}$ is the
number of isotherms, $N_{\mr{P}}(j)$ is the number of experimental data taken
at $T_j$ temperature, $N_{\mr{T}}$ is the total number of experimental data,
$q(P_i,T_j)$ is the calculated saturation, and $q_{\mr{exp}}(P_i,T_j)$ is the
experimental saturation. It can be noticed in Table \ref{tbl:tab1} that
the deviation parameter ($D$) in all cases is less than $5\%$.

In Fig. \ref{fig:fig1} we fit the experimental isotherms of $\mr{SF_6}$ adsorption
on pillared clay (designated as W-A(673)) measured by \citet{bandosz} 
with Eq. (\ref{eq:isomix})\footnote{For
this system, the saturation is expressed as a volume at 101.325 kPa and 273.15
K}. It can be noticed a good agreement between
the experimental data and Eq. (\ref{eq:isomix}). For this set of isotherms, the
correction term (Eq. (\ref{eq:icr1})) was used ($l=1$) and it was obtained a
deviation of 2.97 \%. If this correction were not applied ($l=0$) the deviation
would be 5.2 \%. 
Fig. \ref{fig:fig1} suggests that Eq. (\ref{eq:isomix}) is useful for isotherms with
complex shapes like those reported in Ref. \cite{bandosz}. These experimental
isotherms were also fitted using the 5-parameter Toth
equation and it was obtained a deviation of 17.8\%. 
On the other hand, Eq. (\ref{eq:isott}) could be
 used to fit these experimental isotherms, but more parameters would be
necessary and this would lead to large error bars and thermodynamic properties
with larger uncertainties. Bandosz et al. \cite{bandosz} reported
additional single-component experimental isotherms of propane and sulfur
hexafluoride adsorption on various heat-treated pillared clays; a total of six
adsorbate+adsorbent systems were studied. Given that the results are similar,
here only two analysed systems are reported in Table \ref{tbl:tab1}. However,
the deviation was within 1 and 3 \%  for the six adsorbate+adsorbent systems.  

Fig. \ref{fig:fig2} shows the estimated isosteric heat of adsorption as a function
of adsorbed volume and the corresponding marginal confidence
intervals \cite{bates,seber} for the $\mr{SF_6}$+W-A(673) system. This Fig.
shows that $q^{\mr{st}}$ is nearly constant around 25.5
kJ/mol. Bandosz et al. \cite{bandosz} used a virial
isotherm \cite{jag1,jag2} to correlate their data. They obtained 
plots of $q^{\mr{st}}$ with some oscillations near zero saturation around
25 kJ/mol; this result does not agree with that obtained here by means of Eq.
(\ref{eq:isosh}). These differences might be due to the method used by
Bandosz et al. \cite{bandosz}. They fitted small subsets of 15 adsorption data by using
the virial-type isotherm and four parameters to fit each subset. Although
this method is appropriate to fit the saturation data, the error bars reported by them correspond to
standard deviations in the calculated isosteric heats of adsorption. If 95 \%
marginal confidence intervals are used, the error bars at low coverage would
be larger and the oscillations in the isosteric heat might be within such error bars.

To illustrate the dependence of the isosteric heat of adsorption in terms of
temperature, plots of $q^{\mr{st}}$ vs $T$ at three different saturations are
shown in Fig. \ref{fig:fig3}. As expected, the isosteric heat of adsorption does
not significantly change within the experimental range of temperatures; this
is typically observed in
both experiments and molecular simulations. In the case in which it is
necessary to take into account the dependency of isosteric heat of adsorption
on temperature, the well known thermochemical formula \cite{prigo} to calculate
chemical equilibrium constants could be used to estimate $\ln K_j$ as a
function of temperature in Eq. (\ref{eq:bb}); an empirical model for the heat
capacity would be necessary.

The fitting of the experimental isotherms of adsorption of
1,1,1,2,3,3,3-heptafluoropropane (HFC-227ea) on activated carbon is shown in Fig.
\ref{fig:fig4}. As in the previous case, the fittings are very accurate; 
the correction term was not necessary ($l=0$). Due to the
complex pore structure of activated carbon, it is difficult to consider a
subsystem as any specific region of the real adsorbent, for this reason it is
convenient to regard each cell as an effective subsystem. The isosteric
heat of adsorption calculated by using both Eq. (\ref{eq:isosh}) and the Toth equation are
shown in Fig. \ref{fig:fig5}. It can be noticed that both curves fairly agree. The
differences are attributed to the fact that Yun et al. \cite{yun} only
used two isotherms to calculate the isosteric heat of adsorption. They
employed the parameters of each isotherm and the Clausius-Clapeyron equation
to obtain the $q^{\mr{st}}$ vs $q$ plot. In contrast, here the complete set of
experimental isotherms was used. A problem with the method of
Yun et al. \cite{yun} is that the isosteric heat of adsorption is influenced by
uncertainties in the adsorption data, and hence, very reliable isotherms are
necessary to determine $q^{\mr{st}}$ if only two isotherms are to be employed.

The adsorption isotherms of $\mr{CO_2}$ on
zeocarbon \cite{lees1} and the fractions of molecules distributed among cells
at 273.15 K and 313.15 K are shown in Fig. \ref{fig:fig6}. The zeocarbon synthesized by
Lee et al. \cite{lees1} is a zeolite X/activated carbon mixture
composed by 38.5 mass \% zeolite X, 35 mass \% activated carbon, 10 mass \%
inert silica, and 16.5 mass \% zeolite A and P. For this system, Eq.
(\ref{eq:isomix}) accurately fits the experimental data ($D=3.8\%$), and the
correction term ($l=1$) in this case reduces the deviation by 2 \% with
respect to the $l=0$ case. The deviation is less than that obtained by using the Toth
equation; which is one of the most used equations to fit these kind of
adsorption data. 
Since the zeocarbon is a zeolite/activated carbon
composite, that the adsorbate phase may be
divided into identical weakly-interacting effective subsystems is again a suitable assumption for
fitting and correlation purposes.

As it was mentioned in the previous section, at low pressures the leading term is $f_1$ and each $f_j$
significantly contributes within a certain region of the isotherm. It can be
noticed in Fig. \ref{fig:fig6}a that the saturation ($q$) changes 
approximately $0.5\,\mr{mol\,kg^{-1}}$ near 100 kPa. At 
the lowest experimental temperature for which at high pressures
the saturation is approximately $q_{\mr{m}}$, the leading term is
$f_{n_{\mr{s}}+1}$, as shown in Fig. \ref{fig:fig6}b. However, as temperature
increases in the high experimental pressure region, the isotherm turns into a
combination of several fractions $f_j$. For example, Fig. \ref{fig:fig6}b shows
that the isotherm at 313.15 K in the high pressure region is a combination of
$f_3,\,f_4$ and $f_5$. When temperature is further increased, the term $f_5$
does not contribute to the isotherm. If the correction term is not used and
$n_{\mr{s}}=4$, then the plot of $f_4$ vs $P$ is quite similar to that shown in
Fig. \ref{fig:fig6}c for $f_5$. Thus, this term does not significantly
contribute to the isotherm at 313.15 K. For this reason it was necessary to
propose the correction terms shown in Eqs. (\ref{eq:crr2}) and (\ref{eq:icr1}).
These Eqs. assure that high-order fractions depend on parameters that
substantially contribute to estimate each isotherm within the experimental
temperature range. Because of the specific characteristics of each
adsorbate+adsorbent system, it is difficult to establish a priori whether the
correction term is necessary; it must be tested whether this correction
reduces the fitting standard deviation.

 The results of the fittings to the data reported in
Ref. \cite{watson} are shown in Table \ref{tbl:tab2}. In this case, Eq. (\ref{eq:isott}) was used
without corrections ($l=0$). For the $\mr{CO_2}$+MSC and $\mr{N_2}$+MSC systems, Eq. 
(\ref{eq:isott}) gives better results (in terms of deviation) than the
simplified model (Eq. (\ref{eq:isomix})). The deviation obtained with Eq. (\ref{eq:isomix})
is around 3.5\% ($l=0$), whereas Eq. (\ref{eq:isot}) gives deviations less than 3
\%. Also, the deviation obtained using Eq. (\ref{eq:isott}) is less than that
obtained by using the Toth equation; this is also confirmed by F-tests. The
confidence intervals for $\ln R_j^{\circ}$ are larger than those obtained for
$\ln K^{\circ}$ in Table \ref{tbl:tab1}. This could be caused by parameter
correlation effects. To fit this set of isotherms,
Watson et al. \cite{watson} used the Toth isotherm and also obtained
large confidence intervals for the $K^{\circ}$ parameter. For the $\mr{CH_4}$
system, the Toth isotherm gives better results than Eqs. (\ref{eq:isot}) and
(\ref{eq:isomix}). This is due to an isotherm at 148 K that present an apparent
saturation limit that is less than each of the apparent saturation limits of the
isotherms at temperatures greater than 148 K. If this isotherm at 148 K is
ignored, then the results obtained by using the 5-parameter Toth Eq. and Eq.
(\ref{eq:isott}) are quite similar.

The advantage of the present models is the flexibility of setting the number
of adjustable parameters; this condition is essential to fit isotherms with
complex shapes. Many of the widely used empirical
isotherms to describe type I isotherms like the Sips, Toth, and Dubinin-Ashtakov
isotherms have fixed number of parameters. Moreover, these Eqs. do not reduce to the
correct Henry's law limit; except the Toth isotherm, but it overestimates
the Henry's constant \cite{smith}. In contrast, the GSTA model
has the advantage that they present the correct Henry's law limit and
thus they can be used to estimate this constant. 
Despite its advantages, the model studied here cannot give
site energy and pore size distribution. Although the model does not
explicitly take into account the adsorbent heterogeneity, the number of
parameters and the degree of the polynomial $\xi$, which are related to the
subsystem size, might offer information about the variety of adsorbent sites and
molecule-molecule interaction.



%% References


\section{Conclusions}
It was found that the Eqs. (\ref{eq:isott}) and (\ref{eq:isomix})
can be applied to fit experimental data of adsorption of gases and vapours on
microporous heterogeneous adsorbents. A simple correction that improve 
the fitting results was proposed.
However, this correction may not be necessary in some cases, it must be
tested. Additionally, for systems in which the experimental temperature range
is large, it is suggested that the dependence of $\Delta h_j$ on temperature
should be considered and a model for both gas and adsorbate phase heat capacity
could be required. The advantages of the
GSTA model are the high accuracy that can be achieved to correlate saturation and
thermodynamic data, the flexibility to set the number of adjustable parameters
and consider variations of $\ln K_j$ with temperature, and the possibility
of regarding the adsorptive as a real gas phase. However, the method studied here
does not explicitly consider the pore size and adsorption site energy distribution,
but the size of a representative subsystem offers an idea of the adsorbent
heterogeneity because the size of the subsystem depends on this factor.

\section{References}

\bibliography{Ref}






% \end{thebibliography}

\newpage

\setlength{\textheight}{720pt} \addtolength{\textwidth}{2.0cm}
\addtolength{\hoffset}{-2.0cm} \addtolength{\topmargin}{-3.0cm}
\linespread{1.0}

{\bfseries Figures:}\\
Figure 1. Comparison between the experimental data of $\mr{SF_6}$ adsorption on W-A(673) \cite{bandosz} 
and Eq. (\ref{eq:isot}); symbols: experiment, solid line: Eq. (\ref{eq:isomix}).\\
Figure 2. Plot of isosteric heat of adsorption vs adsorbed volume, the system temperature is 283.0 K;
error bars correspond to 95 \% confidence intervals.\\
Figure 3. Isosteric heat of adsorption for $\mr{SF_6}$+WA-(673) as a function of temperature at three different saturations.\\
Figure 4. Comparison between the experimental data of HFC-227ea adsorption on activated carbon \cite{yun} 
and Eq. (\ref{eq:isomix}); symbols: experiment, solid line: Eq. (\ref{eq:isomix}).\\
Figure 5. Calculated isosteric heat of HFC-227ea adsorption on activated carbon; dotted/dashed line: calculated by Yun et al. \cite{yun}, solid line: Eq. (\ref{eq:isosh}), dotted lines: 95 \% marginal confidence intervals for Eq. (\ref{eq:isosh}).\\
Figure 6. a) Comparison between the experimental data of $\mr{CO_2}$ adsorption on zeocarbon \cite{lees1} 
and Eq. (\ref{eq:isot}); b) distribution of molecules among cells at 273.15 K; 
c) same as b) for 313.15 K.

\begin{figure}[H]
\centering
\includegraphics[width=1.2\textwidth]{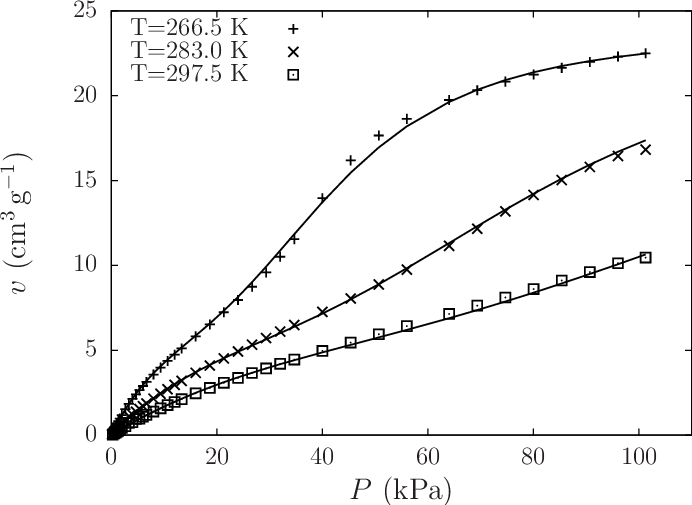}
\caption{}
\label{fig:fig1}
\end{figure}

\begin{figure}
\centering
\includegraphics[width=1.2\textwidth]{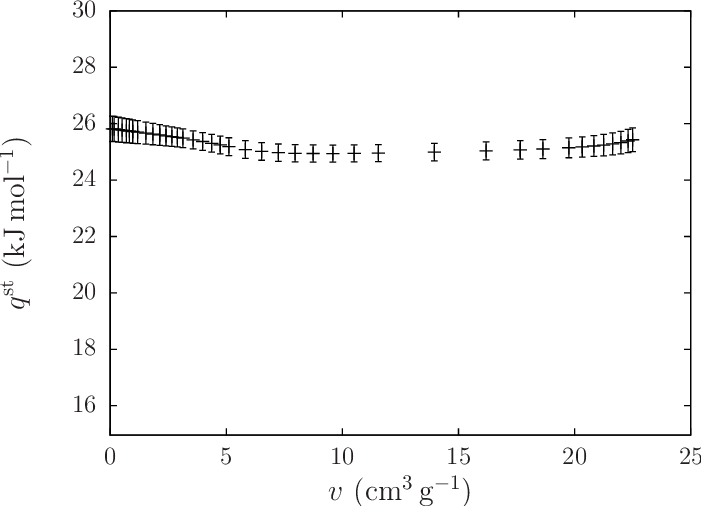}
\caption{}
\label{fig:fig2}
\end{figure}

\begin{figure}
\centering
\includegraphics[width=1.2\textwidth]{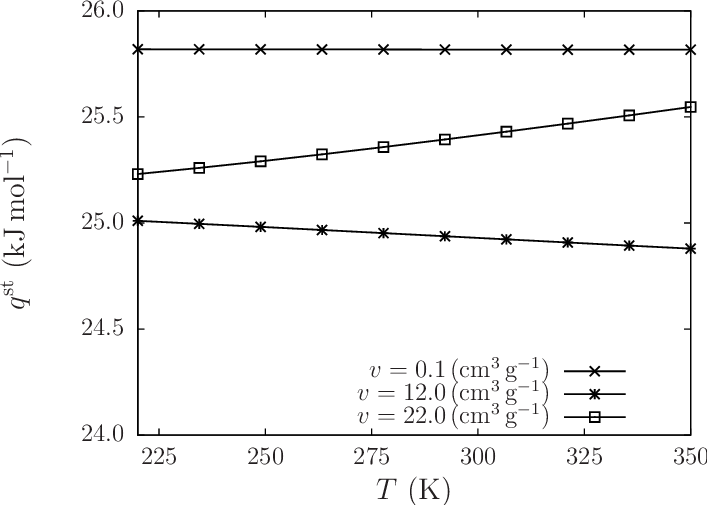}
\caption{}
\label{fig:fig3}
\end{figure}

\begin{figure}
\centering
\includegraphics[width=1.2\textwidth]{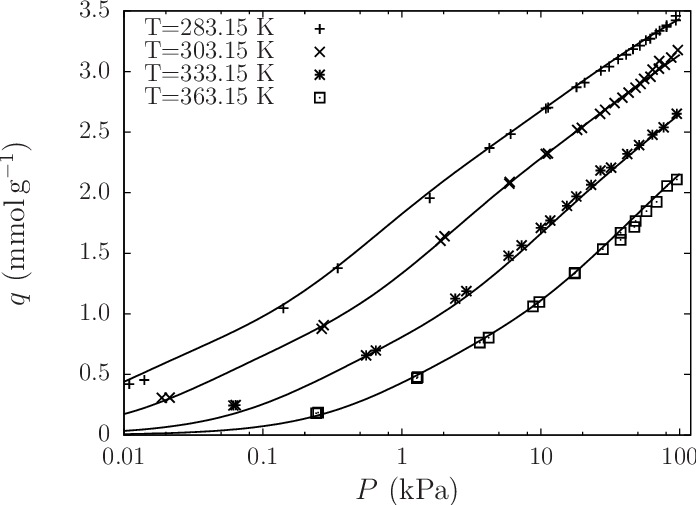}
\caption{}
\label{fig:fig4}
\end{figure}

\begin{figure}
\centering
\includegraphics[width=1.2\textwidth]{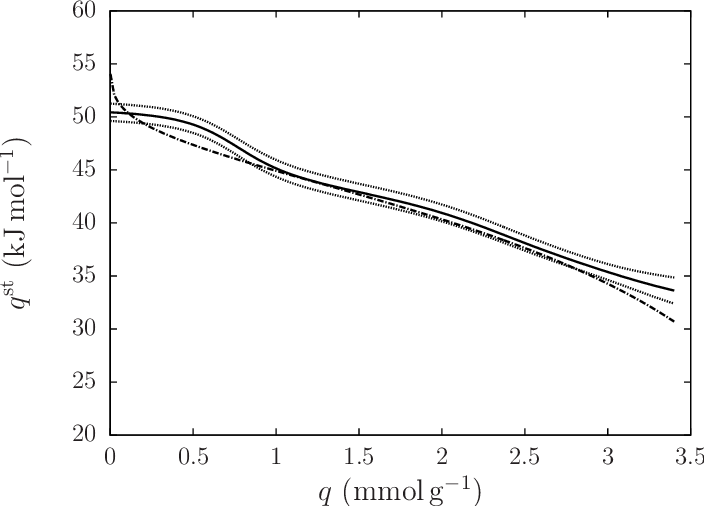}
\caption{}
\label{fig:fig5}
\end{figure}

\begin{figure}
\centering

\includegraphics[width=0.6\textwidth]{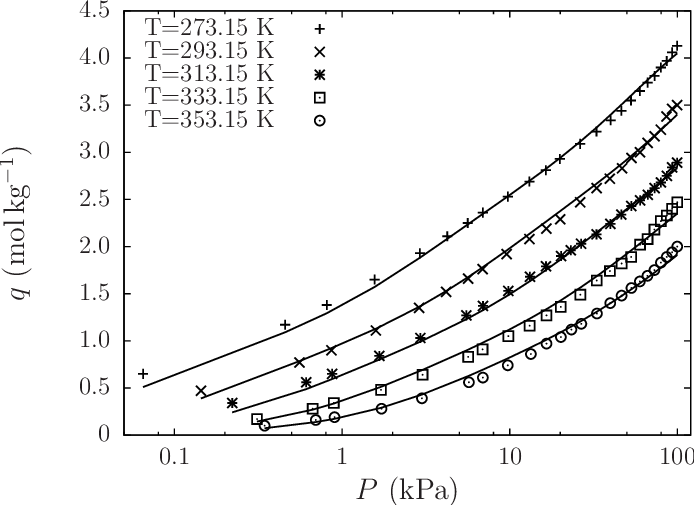}\\
a)

\includegraphics[width=0.6\textwidth]{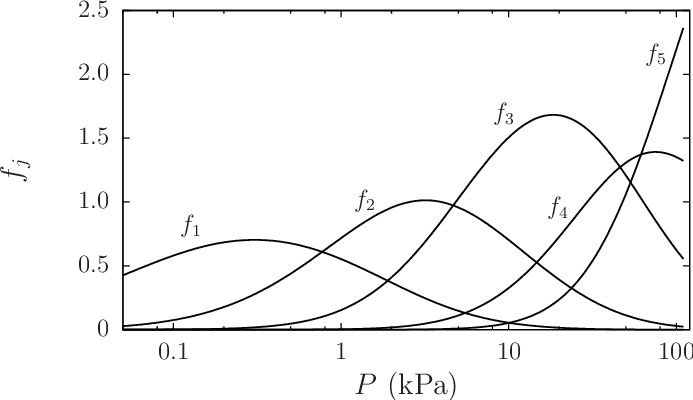}\\
b)

\includegraphics[width=0.6\textwidth]{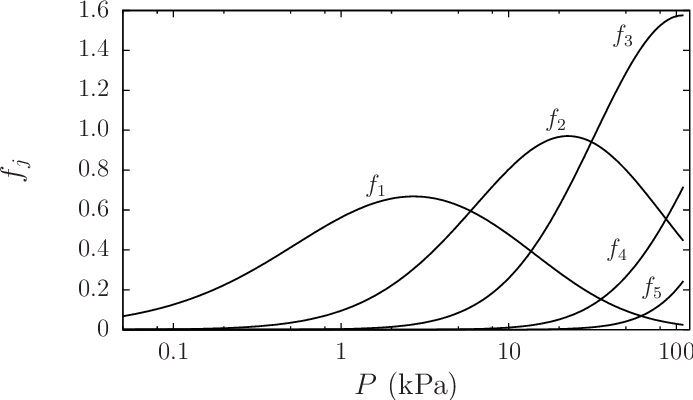}\\
c)

\caption{}
\label{fig:fig6}
\end{figure}

\begin{table}[H]
\begin{center}
\caption{Estimated parameters for several systems using Eq. \ref{eq:isomix}.\textsuperscript{\emph{a,b,c,d}}}
\label{tbl:tab1}
\begin{footnotesize}
\begin{tabular}{ccccccccc}                                        %%%%%%CORRECCIÓN
       & $q_{\mr{m}}$ (or $v$) & $\ln K^{\circ}$ &    $-10^{-3}\times \Delta h_j/R\, $                  & $D$  & Temperature & Pressure\\
System & $\mr{mmol\,g^{-1}}$ (or $\mr{cm^3\,g^{-1}}$)  &   &                     $\mr{K}$               &    $\%$  & range (K) & range (kPa)   \\
\hline

$\mr{SF_6}$+W-A(673)\cite{bandosz}    & $23.89\pm 0.44$ & $-13.67\pm 0.16$ & $3.107\pm 0.055$ & 2.97 & 266.5-297.5 &  0.1-100\\
      $l=1$                         &                &                   & $6.112\pm 0.077$ &      &   &\\
                                    &                &                   & $8.08\pm 0.50$ &       &    &\\
                                    &                &                   & $11.62\pm 0.32$ &      &   &\\

$\mr{C_3H_8}$+W-A\cite{bandosz}        &$34.6\pm 3.9$ &$-14.65\pm 0.30$ & $3.391\pm 0.096$ & 3.2 & 267-298 & 0.1-100 \\    %    3.24 sin la corrección         
            $l=1$                   &                &                 & $7.472\pm 0.097$ &      && \\
                                    &                &                 & $10.8\pm 0.23$&       && \\
                                    &                &                 & $14.8\pm 0.60$&      && \\
                                    &                &                 & $17.9\pm 0.48$&      &&\\

$\mr{HFC}$-227ea+AC\cite{yun}         &$3.87\pm 0.27$  &$-16.67\pm 0.27$          & $6.066\pm 0.098$  & 2.36 & 283.15-363.15 & 0.01-100\\                  
              $l=0$                 &                &                 & $11.39\pm 0.19$  &     &&  \\
                                    &                &                 & $16.38\pm 0.32$  &    && \\
                                    &                &                 & $20.82\pm 0.45$  &    && \\
                                    &                &                 & $24.71\pm 0.68$  &    && \\  
                                    
$\mr{HFP}$+AC\cite{yun}               &$3.87\pm 0.20$  &$-15.66\pm 0.20$ & $5.429\pm 0.075$ & 2.53 & 283.15-363.15 & 0.01-100 \\                  
              $l=0$                      &                &                 & $10.15\pm 0.15$  &      && \\
                                    &                &                 & $14.68\pm 0.23$  &     && \\
                                    &                &                 & $18.69\pm 0.32$  &     && \\
                                    &                &                 & $22.33\pm 0.45$  &     && \\
                                    
$\mr{CO_2}$+ZC\cite{lees1}             &$4.82\pm 0.39$  &$-15.71\pm 0.29$ & $5.03\pm 0.10$ & 3.84  & 273.15-353.15 & 0.05-100\\                  
               $l=1$                &                &                 & $9.39\pm 0.20$  &      && \\
                                    &                &                 & $13.45\pm 0.31$  &     && \\
                                    &                &                 & $17.00\pm 0.43$  &     && \\  
                                    
$\mr{CO_2}$+SG\cite{berlier}           &$14.5\pm 1.0$&$-15.22\pm 0.10$ & $3.068\pm 0.030$ & 1.47 & 278-328 & 50-3400  \\%%%%%Aquí se usó YYC                  
             $l=4$                  &                &                 & $5.759\pm 0.052$  &    &&   \\
                                    &                &                 & $8.454\pm 0.094$  &   &&  \\                                   
                                    &                &                 & $10.87\pm 0.12$  &   &&

\end{tabular}
\end{footnotesize}
\end{center}

\begin{flushleft}
\begin{footnotesize}
\textsuperscript{\emph{a}} In all cases, $P^0$ is expressed in kPa.\\
\textsuperscript{\emph{b}} Abbreviations: activated carbon (AC), 1,1,1,2,3,3,3-heptafluoropropane (HFC-227a), 
hexafluoropropene (HFP), zeocarbon (ZC), silica gel (SG).\\
\textsuperscript{\emph{c}} The parameters $\Delta h_j$ are tabulated in increasing order of $j$, e.g., 
 for the $\mr{SF_6}$+W-A(673) system, $\Delta h_1/R=-3.107\times 10^3,\,\Delta h_2/R=-6.112\times 10^3$,
 and so on.\\
%\textsuperscript{\emph{d}} For systems \ce{SF6}+MSC, \ce{C3H8}+MSC, and \ce{CO2}+ZC, eq \ref{eq:icr1}
%was applied. For system \ce{CO2}+SG, Eqs. \ref{eq:icr1} and \ref{eq:icr2} were applied. For systems
%HFC+AC and HFP+AC, no corrections were applied.\\
\textsuperscript{\emph{d}} For $\mr{SF_6}$+W-A(673) and $\mr{C_3H_8}$+W-A systems, the saturation 
is expressed in $\mr{cm^3\,g^{-1}}$ at 101.325 kPa and 273.15 K, and for the remaining systems it is
expressed in $\mr{mmol\, g^{-1}}$. 
\end{footnotesize}
\end{flushleft}
\end{table}

\begin{table}[H]
\begin{center}
\caption{Estimated parameters for the adsorption data presented by Watson et al.\cite{watson} using Eq.
\ref{eq:isott} (l=0).\textsuperscript{\emph{a,b,c}}}
\label{tbl:tab2}
\begin{footnotesize}
\begin{tabular}{ccccccccc}
       & $q_{\mr{m}}$ & $\ln R_j^{\circ}$ & $-10^{-3}\times\Delta\overline{h}_j/R\, $  & $D$  & Temperature & Pressure\\
System & $\mr{mmol\cdot g^{-1}}$ &   &                     $\mr{K}$               &    $\%$   &range (K) & range (kPa)   \\
\hline
$\mr{CH_4}$+MSC                        & $4.412\pm 0.067$ & $-14.4\pm 1.5$ & $2.81\pm 0.41$ & 2.29 & 148-298 & 1-4000 \\
                                    &                &$-13.88\pm 0.48$& $1.92\pm 0.13$ &       &            &           \\
                                  
$\mr{CO_2}$+MSC                        & $5.87\pm 0.14$ & $-13.7\pm 1.3$ & $3.26\pm 0.40  $ & 1.41 & 223-323 & 25-5200 \\
                                    &                & $-14.00 \pm 0.83$ & $2.94\pm 0.25$ &       &&\\
                                    &                & $-16.07\pm 0.82$ & $3.29\pm 0.22$ &       &&\\
                                    &                & $-19.46\pm 0.58$ & $3.58\pm 0.17$ &       &&\\
                                  
$\mr{N_2}$+MSC                         & $5.75\pm 0.12$ & $-15.6\pm 1.1$ & $2.71\pm 0.31$ & 3.14  & 115-298 &  0.01-5000 \\                  
                                    &                &$-14.03\pm 0.45$& $1.725\pm 0.099$ &       &&\\
                                    &                & $-18.3\pm 1.0$ & $1.79\pm 0.15$ &       &&\\
                                 
\end{tabular}
\end{footnotesize}
\end{center}

\begin{flushleft}
\begin{footnotesize}
\textsuperscript{\emph{a}} In all cases, $P^0$ is expressed in kPa.\\
\textsuperscript{\emph{b}} Abbreviation: molecular sieving carbon (MSC).\\
\textsuperscript{\emph{c}} As in \ref{tbl:tab1}, the parameters $\ln R_j^{\circ}$ and $\Delta \overline{h}_j$
 are tabulated in increasing order of $j$, for example, for the $\mr{CH_4}$+MSC system, 
 $\ln R_1^{\circ}=-14.4,\,\ln R_2=-13.88$, $\Delta \overline{h}_1/R=-2.81\times 10^3,\,\Delta \overline{h}_2/R=-1.92\times 10^3$.
\end{footnotesize}
\end{flushleft}
\end{table}

\end{document}